# AN UNBALANCED FEED DESIGN FOR WIDEBAND PHASED ARRAYS

David W. Landgren, Daniel J. P. Dykes, and Kenneth W. Allen
Advanced Concepts Laboratory, Georgia Tech Research Institute,
Georgia Institute of Technology, 30332 Atlanta, GA, USA
Corresponding author: kenneth.allen@gtri.gatech.edu

## ABSTRACT

In this work, a planar phased array antenna was engineered with ultra-wideband (UWB) performance that covers portions of the L-, S-, and C-bands. The unit cell design contains a driven layer that is directly connected to a single coaxial feed and a parasitic layer located above the driven layer separated by free space. This design does not require a balun or any vias between the antenna ground plane and the driven layer, resulting in a simple antenna stack-up consisting of only planar layers, aside from the feed structure. As a consequence, the complexity, and potentially sensitivity to mechanical tolerances, is reduced. The simulated results of this unbalanced phased array are discussed and experimentally validated.

## INTRODUCTION

Recently, the government auctioned off sections of the electromagnetic spectrum to industry and other entities. This has resulted in a transition of certain government systems from L-Band to higher frequencies such as C-Band. In this work, an UWB antenna system was designed to support fire-control radar testing that will instantaneously cover both the legacy L-Band frequencies and the new C-Band frequencies. Additionally, the planar array has a reduced form factor (thickness $< \lambda/10$ at the highest frequency of operation) compared to existing, more traditional solutions, e.g. parabolic dish antenna. The key to minimizing antenna thickness is designing the array without baluns or commercial-off-the-shelf (COTS) combiner networks; this proceeding will focus on a technique to design arrays without baluns.

Traditionally, wideband planar arrays are designed with a differential feeding scheme. That is, each element (or unit cell) of the phased array is fed with a pair of transmission lines that are ideally supplied with signals that are equal in amplitude and have a 180-degree phase difference. Assuming the aperture geometry is also symmetric, this feeding scheme results in the flow of electrical current in a preferred direction across the unit cell, a common technique for producing a linear polarization state. Furthermore, this feature of the electrical current supports UWBs from planar arrays as described by the well-known current sheet concept developed by Wheeler [1].

Phased arrays with unbalanced feeding schemes simplify the front-end antenna system since no baluns are required; however, these antenna arrays have resulted in more complex architectures for the unit cell. In some cases, this has required vias to be positioned in very close spatial proximity to the feed structure [2-7]. The spatial proximity leads to tight electromagnetic coupling between the vias and feed structure; consequently, the antenna performance of such designs can be sensitive to mechanical deviations from the nominal design values. In this work, a more stable solution was engineered by removing the vias altogether. It will be shown that this can be done without a significant reduction in performance, as the realized gain of a fully driven array is within 1.5 dB of the area gain limit for arrays larger than 8x8.

The remainder of this work is organized as follows. The first section will provide details of the unit cell geometry. The second section will present two designs: (*i*) a single-layer design and (*ii*) a two-layer design, comprised of a driven layer and a parasitic director layer. The third section will provide a discussion of how performance is affected by array size and options for implementing a power combiner. Finally, a conclusion will summarize this work and project how this technology can be utilized in the future.

## PHASED ARRAY UNIT CELL GEOMETRY

The antenna unit cell consists of multiple layers of low-permittivity ($\varepsilon_r$) materials, a standard 0.085" coaxial feed, and multiple patterned copper layers (more specifically fragmented aperture layers). The specific materials are a low-density General Plastics foam FR-4540 ($\varepsilon_r = 2.0$) and Rogers 5880LZ ($\varepsilon_r = 1.96$). The low-density foam provides mechanical support and separation between the aperture and the ground plane with the desirable properties of light weight, machinability, and minimum electromagnetic interaction due to the low-permittivity. The two Rogers 5880LZ boards have 0.5 oz copper that is patterned for the radiating layer and continuous for the ground plane. An example stack-up is shown in Figure 1, illustrating a single feed and no additional conductors or vias. This is an indication that no baluns are required.

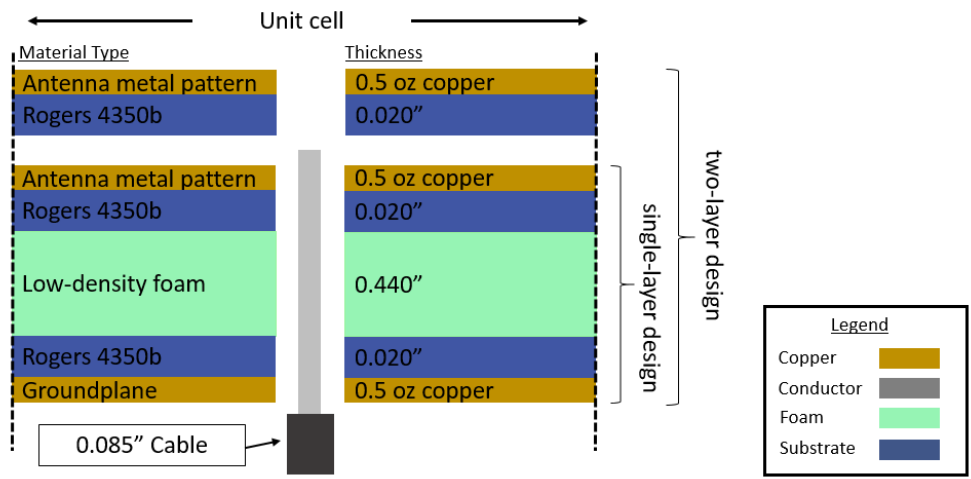

Figure 1. Antenna material stack-up of the array unit cell.

The exact details of the fragmented layer are determined by a process developed by the Georgia Tech Research Institute (GTRI) over the last decade and a half [8]. This process leverages an evolutionary algorithm to design antennas; however, this process can be extended to evaluate, characterize, and generate other electromagnetic structures [9-17]. The general form factor is chosen based on physical principles (e.g., area gain limit, avoiding grating lobes, ground plane spacing, etc.) and application-specific features (e.g., feed structure, restricted footprint, etc.). The aperture is integrated within this form factor, Figure 1, and is a fragmented layer consisting of discrete areas, Figure 2, that can be toggled between conducting and dielectric by a genetic algorithm to optimize the performance of the antenna. A full 3D finite-difference time-domain (FDTD) electromagnetic simulation is conducted for each candidate antenna solution, and the performance is compared against some predefined goal (e.g., gain vs. frequency). In the case of this work, the goal was to maximize realized gain vs. frequency from L-band to C-band.

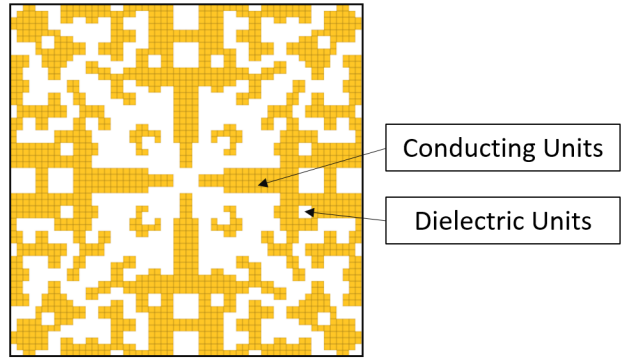

Figure 2. Example fragmented layer.

## OPTIMIZED DESIGNS OF THE PHASED ARRAY UNIT CELL

A unit cell was designed with a single fragmented radiating layer (i.e., driven layer). This layer is directly excited by the center conductor of a coaxial feed. The simulated (using an in-house FDTD code) realized gain, reflection coefficient, and complex input impedance of this design is shown Figure 3. The realized gain has been normalized to 0 dBiL, corresponding to 100% efficiency. This design has two minima in the gain across the optimized band, at approximately 2.5 and 5.25 GHz. Inspecting the input impedance at these two frequencies reveals that the antenna is more capacitive than the rest of the operating band. This insight led to the idea of adding an additional layer for tuning to help improve the complex input impedance, where the imaginary component can be dominated by capacitance or inductance, to provide a flattened resistance and reactance profile over the frequency bands of interest.

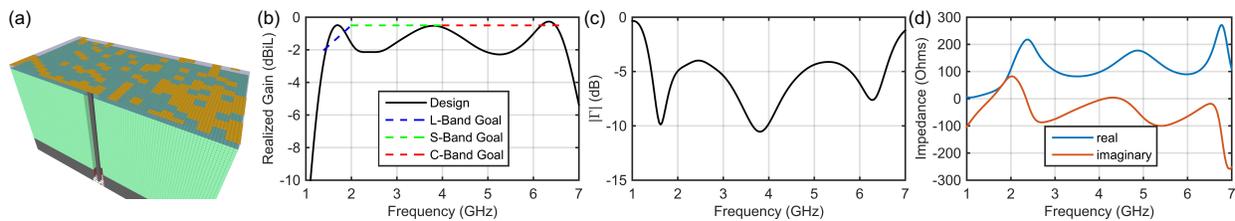

Figure 3. Single-layer antenna design. (a) An image showing half of the unit cell, (b) simulated results of the infinite array gain, (c) reflection coefficient, and (d) complex input impedance as a function of frequency.

In order to address the degradation in gain of the single-layer design, a two-layer antenna was designed. The second layer acts as a parasitic tuning mechanism that resides at some distance (i.e., 7.5 mm) away from the driven layer, determined by the optimizer. During this optimization, the fragmented pattern of the driven layer was held fixed to match the single-layer design, Figure 3(a). The predicted performance is shown in Figure 4. The additional layer has helped increase the infinite array gain by approximately 0.5 dBiL, an appreciable improvement over the previous design. Given that all of the materials for this design are low-loss dielectrics ($tan\delta < 0.01$), the majority of the improvement is associated with the impedance match. The impedance plot in Figure 4(d) confirms this, given that both components of the complex impedance are flatter and have reduced extrema across the operating band.

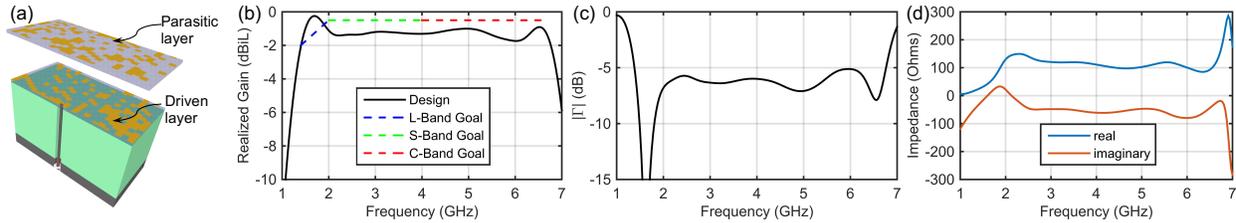

Figure 4. Two-layer antenna design. (a) An image showing half of the unit cell, (b) simulated results of the infinite array realized gain, (b) reflection coefficient, and (d) input impedance as a function of frequency.

## EXPERIMENTAL RESULTS

A small 4x4 array prototype was built and measured based on the single-layer design. An image of this antenna array, along with an inset of the unit cell, is shown in Figure 5(a). The array is fully cabled, i.e. each element is directly connected to a cable, so that each element can be characterized individually. In order to mechanically support the 16 cables and the antenna, a fixture was built with plastic posts and a metal plate, as shown in Figure 5(a). It should be noted that the fixture is electromagnetically shielded by the antenna ground plane and should have minimal impact on the antenna performance. The total gain is calculated by performing a complex sum of all 16 ports of the 4x4 array using uniform amplitude weighting. This mimics a perfect power combiner and therefore is ideal. The experimental results of the measured realized gain agree with the simulation, as shown in Figure 5(b). For telemetry applications, a combiner network would be integrated with the antenna array, resulting in a single RF-connector output. However, the development of the combiner network goes beyond the scope of this proceeding.

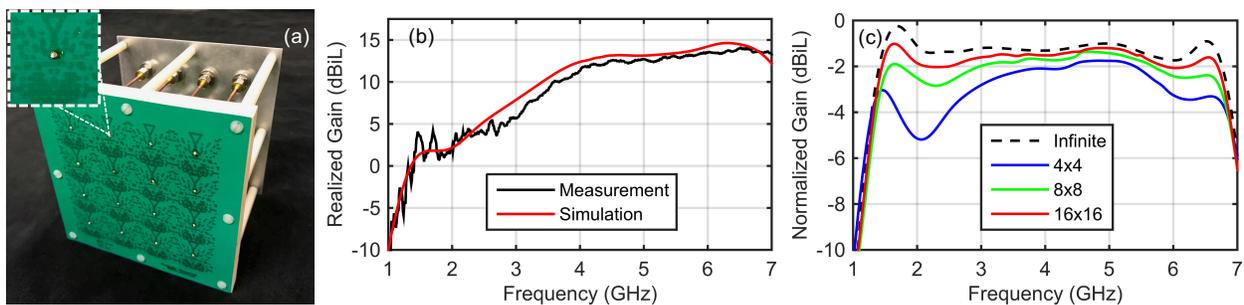

Figure 5. Small prototype antenna array. (a) image of the assembled array with an inset of a single unit cell, (b) measured broadside gain compared with the simulated result, and (c) simulated results for varying area sizes compared with an infinite array in terms of normalized gain.

## DISCUSSIONS

Ultimately, a larger array (tentatively 16x16) will be fabricated in order to support the gain requirements for specific test ranges. Smaller arrays can significantly deviate from performance of the infinite case due to array truncation effects. These deleterious effects are especially present at the lower frequencies, where the antenna is less than a wavelength in some cases. Generally, as the array becomes larger, the performance converges towards the infinite array result. This is illustrated in Figure 5(c), which shows the trend in performance for various sized arrays: 4x4, 8x8, and 16x16. The 16x16 is nearly converged to the results from the infinite array result, i.e. periodic boundary conditions applied in the plane of the unit cell. This helps build confidence that the array designed in the infinite array setting will perform as anticipated, if tiled to 16x16 or greater. It should also be noted that the experimental results presented are for the single-layer design without the tuning layer. Prototyping of the two-layer design will be the subject of future work.

COTS products are available to support a 16:1 power combiner network. However, in-house combiners are also being designed, leveraging the same design procedure used for the fragmented antenna aperture. These fragmented combiners are part of an initiative to explore the more general space of fragmented circuitry. Fragmented combiner networks could potentially provide superior performance as the combiner network could be optimized simultaneously with the antenna aperture, tuning for deviations from the assumed 50 $\Omega$ load.

## CONCLUSIONS

In this work, a planar phased array antenna was demonstrated with UWB performance that covers portions of the L-, S-, and C-bands with FDTD simulations, validated by experimental measurements. The unit cell design is simplified in comparison to the planar unbalanced modular arrays (PUMA) in refs. [2-7] and offers a new approach to this problem. The unit cell includes a DC-connected driven layer with a parasitic layer above to tune the input impedance of the array. This simple stack-up does not require a balun or vias from the driven layer to the ground plane. Consequently, this simple antenna stack-up consists of only planar layers, aside from the feed structure. Therefore, the structural complexity is reduced which could lead to a more robust solution in terms of sensitivity to mechanical deviations. The concepts developed in this work are flexible and a suitable solution for scaling to higher frequencies, seeing as the structures are mainly dependent on the spatial resolution of the planar features. Regarding intended telemetry applications, the antenna will be integrated with an appropriate combiner network resulting in a single RF feed for interfacing with test range systems.

## ACKNOWLEGEMENTS

This project is managed by the Test Resource Management Center (TRMC) and funded through the Spectrum Access R&D program via Picatinny Arsenal under Contract No. WL5QKN-15-9-1004. The Executing Agent and Program Manager work out of the AFTC. Any opinions, findings, and conclusions or recommendations expressed in this material are those of the authors and do not necessarily reflect the views of TRMC, the SAR&D Program and/or the Picatinny Arsenal.

# REFERENCES

[1] Wheeler, Harold. "Simple relations derived from a phased-array antenna made of an infinite current sheet." *IEEE Transactions on Antennas and Propagation* 13.4 (1965): 506-514.
[2] Holland, Steven S., and Marinos N. Vouvakis. "Design and fabrication of low-cost PUMA arrays." *Antennas and Propagation (APSURSI), 2011 IEEE International Symposium on*. IEEE, 2011.
[3] Logan, John T., and Marinos N. Vouvakis. "Planar ultrawideband modular antenna (PUMA) arrays scalable to mm-waves." *Antennas and Propagation Society International Symposium (APSURSI), 2013 IEEE*. IEEE, 2013.
[4] Lee, Michael Y., et al. "Simplified design of 6:1 PUMA arrays." *Antennas and Propagation & USNC/URSI National Radio Science Meeting, 2015 IEEE International Symposium on*. IEEE, 2015.
[5] Holland, Steven S., Daniel H. Schaubert, and Marinos N. Vouvakis. "A 7–21 GHz dual-polarized planar ultrawideband modular antenna (PUMA) array." *IEEE Transactions on Antennas and Propagation* 60.10 (2012): 4589-4600.
[6] Holland, Steven S., and Marinos N. Vouvakis. "The planar ultrawideband modular antenna (PUMA) array." *IEEE Transactions on Antennas and Propagation* 60.1 (2012): 130-140.
[7] Logan, John T., et al. "A review of planar ultrawideband modular antenna (PUMA) arrays." *Electromagnetic Theory (EMTS), Proceedings of 2013 URSI International Symposium on*. IEEE, 2013.
[8] Maloney, James Geoffrey, et al. "Fragmented aperture antennas and broadband antenna ground planes." U.S. Patent No. 6,323,809. 27 Nov. 2001.
[9] Allen, Kenneth W., et al. "An X-band waveguide measurement technique for the accurate characterization of materials with low dielectric loss permittivity." *Review of Scientific Instruments* 87.5 (2016): 054703.
[10] Scott, Mark M., et al. "Permittivity and permeability determination for high index specimens using partially filled shorted rectangular waveguides." *Microwave and Optical Technology Letters* 58.6 (2016): 1298-1301.
[11] Allen, Kenneth W., et al. "Metasurfaces guided by evolution: Ultra-wideband Ku-bandpass frequency selectivity is fit for survival," Submitted (2017).
[12] Westafer, Ryan S., et al. "Micropatterned W-band Antenna Tiles," *IMS 2017 - IEEE MTT-S INTERNATIONAL MICROWAVE SYMPOSIUM,* June 4-7 (2017).
[13] Rahmat-Samii, Yahya, and Eric Michielssen. "Electromagnetic optimization by genetic algorithms." *Microwave Journal* 42.11 (1999): 232-232.
[14] Weile, Daniel S., and Eric Michielssen. "Genetic algorithm optimization applied to electromagnetics: A review." *IEEE Transactions on Antennas and Propagation* 45.3 (1997): 343-353.
[15] Johnson, J. Michael, and V. Rahmat-Samii. "Genetic algorithms in engineering electromagnetics." *IEEE Antennas and Propagation Magazine* 39.4 (1997): 7-21.
[16] Haupt, Randy L., and Douglas H. Werner. *Genetic algorithms in electromagnetics*. John Wiley & Sons, 2007.
[17] Nagar, Jogender, and Douglas H. Werner. "A Comparison of Three Uniquely Different State of the Art and Two Classical Multiobjective Optimization Algorithms as Applied to Electromagnetics." *IEEE Transactions on Antennas and Propagation* 65.3 (2017): 1267-1280.